\documentclass[pre,aps,groupaddress,showpacs,prd,twocolumn]{revtex4}
\usepackage{amsmath}
\usepackage{epstopdf}
\usepackage{color}

\usepackage{epsfig}
\usepackage{graphicx}
 
\usepackage{amssymb}
\usepackage{amsfonts}
\usepackage{dcolumn}
 
\newcommand{\br}{{\bf r}} 
\newcommand{\bv}{{\bf v}} 
\newcommand{\rT}{{\mbox{{\footnotesize T}}}} 

\begin{document}

%\twocolumn[

\title{Two-dimensional dissipative solitons supported by localized gain}

\author{Yaroslav V.  Kartashov$^1$, Vladimir V. Konotop$^2$, and Victor A. Vysloukh$^1$}

\address{$^1$ICFO-Institut de Ciencies Fotoniques, and Universitat Politecnica de Catalunya, Mediterranean Technology Park, 08860 Castelldefels (Barcelona), Spain
 \\
$^2$Centro de F\'{\i}sica Te\'orica e Computacional
  and Departamento de F\'{\i}sica, Faculdade de Ci\^encias, Universidade de Lisboa,
Avenida Professor Gama
Pinto 2, Lisboa 1649-003, Portugal
%\\
%$^3$Departamento de Fisica y Matematicas, Universidad de las Americas - Puebla, Santa Catarina %Martir, 72820, Puebla, Mexico
}

\begin{abstract}
%\par

We show that the balance between localized gain and nonlinear cubic dissipation in the two-dimensional nonlinear Schr\"odinger equation allows for existence of stable two-dimensional localized modes which we identify as solitons. Such modes exist only when the gain is strong enough and the energy flow exceeds certain threshold value. The observed solitons neither undergo diffractive spreading nor collapse. Above the critical value of the gain  the symmetry breaking occurs and asymmetric dissipative solitons emerge. 

\end{abstract}

%\ocis{190.0190, 190.6135}
%]

\maketitle

The nonlinear Schr\"odinger (NLS) equation is probably the most widely used mathematical model describing numerous nonlinear phenomena in the optics of Kerr media~\cite{optics}.
 Being supplied by gain and/or dissipation it is employed for modeling open systems and plays a particularly important role for description of optical patterns~\cite{opt_patt}.  
Soliton formation  described by the NLS equation and soliton properties depend dramatically on the dimensionality of the system. In particular, in the conservative case, while the  one-dimensional (1D) cubic NLS equation with homogeneous coefficients has the well known  soliton solutions, no stable localized states can exist in two or more dimensions~\cite{Segev}. When homogeneous linear gain and nonlinear dissipation are added to the system, stable stationary localized structures do not exist even in the 1D model~\cite{Malomed2}.  Stabilization, however, can be achieved in 1D case by considering a spatially localized gain~\cite{Malomed}.

In this Letter we demonstrate theoretically and numerically that the use of the localized gain source in the presence of nonlinear losses allows one to obtain stable  dissipative solitons also in a 2D dissipative NLS equation.
We consider the model
\begin{eqnarray}
iq_\xi = - \frac 12 \nabla^2 q    +i \gamma(r)q-  |q|^2q - i\alpha |q|^2q
\label{NLS}
\end{eqnarray}
where $q\equiv q(\xi,\br) $ is a complex field, $\xi$ is the propagation variable, $\br=(\eta,\zeta)$ is a  vector in the transverse plane, 
$\nabla\equiv\left(\frac{\partial}{\partial\eta},\frac{\partial}{\partial\zeta}\right)$, the localized gain  $\gamma(r)$ is a real decaying function of $r=|\br|$. It will be convenient to represent $\gamma\equiv p_if(r)$, where $p_i$ is   the gain coefficient and $f(r)$ is a real decaying function with   $\max\{ f(r)\}=f(0)=1$. The  positive constant $\alpha $ characterizes nonlinear losses.

 We notice  recently increasing interest in exploring  localized gain~\cite{Malomed,Kutz,KKVT}, which can be   implemented in various physical systems, like  layered structures pumped by inhomogeneous currents~\cite{Kutz}, nonlinear materials with spatially localized doping~\cite{KKVT}, laser systems~\cite{laser} and fluid mechanics~\cite{fluid}.      
As to Eq.~(\ref{NLS}),  it describes nonlinear response of semiconductor alloys where soliton formation is observable for wavelengths below the half bandgap and the two-photon absorption is the dominating mechanism of losses at power levels required for soliton formation~\cite{solit_form}. Such materials are widely used for production of wide-band semiconductor optical amplifiers with high optical gain in the same spectral range \cite{amplifier}.

 Solutions  of Eq. (\ref{NLS}) predicting dissipative solitons are searched in the form $\psi=w(\br)e^{ib\xi}$, where $b$ is the propagation constant, $w(\br)=w_r+iw_i=u(\br)e^{i\theta(\br)}$ is a complex amplitude, whose real and imaginary parts are respectively $w_r$ and $w_i$, while $u(\br)$ and $\theta(\br)$ are respectively the real amplitude and phase  satisfying the equations   
\begin{subequations}
\label{sys}
\begin{eqnarray}
\label{sys2}
-2bu + \nabla^2 u  -  u \bv^2   + 2 u^3 =0,
\\ 
\nabla\cdot(u^2\bv)+2\gamma(r) u^2 -2\alpha u^4=0,
\label{sys3}
\end{eqnarray}
\end{subequations}
where $\bv=\nabla\theta$. We are interested in the localized solutions with zero asymptotics: $ u, |v| \to 0$  at $r\to\infty$.  

First, concentrating on  solutions with $\bv=0$,
 we show that by controlling the gain one can construct localized modes of Eqs.~(\ref{sys}). 
More specifically,  we pose the problem of finding the gain corresponding to the {\it a priori} given soliton. In the 1D case ($\br=\eta$ and $\nabla=\partial /\partial \eta $) for $v=0$   we find  that  
$ w_{ds}= e^{i/(2d^2)}/\left[d\cosh(\eta/d)\right]$ 
%(having the form of the NLS soliton) 
solves Eqs.~(\ref{sys}) provided the gain  is given by $\gamma(\eta)\equiv (\alpha/ d^2)/ \cosh^2(\eta/d)$, where  $d>0$ is a  constant defining  the localization of the gain. 
Turning, to the 2D case we observe that  for $\bv=0$ Eq.~(\ref{sys2}) has the well known 
%radially symmetric 
Townes soliton solution. We denote it as $w_{\rT}(r)$. In order to support the existence of the {\em dissipative Townes soliton}, the  gain must have the form $\gamma_{\rT}(r)=\alpha w_{\rT}^2(r)$. For any $\gamma(r)\neq \gamma_\rT(r)$ dissipative solitons with a trivial phase do not exist.  
 
In general, i.e. when ${\bf v}$ is not necessarily  zero, stationary localized solutions  exist only for $b>0$; it follows from the limit $r\to\infty$, where one can neglect both nonlinear terms (i.e. $v^2$ and $u^3$) in Eq.~(\ref{sys2}). 
The balance between gain and dissipation is expressed by  
\begin{eqnarray}
\label{balance1}
\int\gamma(r)| w|^2d\br=\alpha\int| w|^4d\br. 
\end{eqnarray}
From (\ref{NLS}) one can also obtain  the equation
\begin{eqnarray}
\label{balance2}
b U+\frac 12 \int |\nabla w|^2 d\br=\int | w|^4d\br
\end{eqnarray}
here $U=\int u^2d\br$ is the energy flow,  that expresses the balance between conservative terms.
 Thus  $b$   is defined by the interplay between gain and dissipation. This imposes restrictions on the range of variation of $b$. Indeed, from (\ref{balance1}) and (\ref{balance2}), combined with the   estimate $\int\gamma  u^2d\br\leq p_iU$, we obtain 
$  \int |\nabla w|^2d\br\leq\left( p_i/\alpha -b\right) U$. Hence  $b\in(0,p_i/\alpha)$. 
   
\begin{figure}[h]
   \begin{tabular}{c} 
\includegraphics[width=0.9\columnwidth]{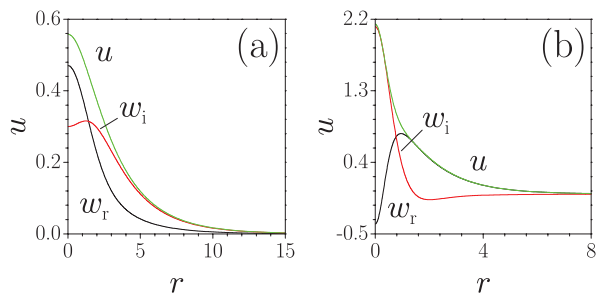}
  \end{tabular}
\caption{(Color online) Profiles of radially symmetric dissipative solitons at (a) $p_i=0.642$, $\alpha=1.2$ and  (b) $p_i=4$, $\alpha=1.9$. Real, imaginary parts of the field, and the field modulus are shown with black, red, and green lines, respectively. In the both cases $d=1.5$.}
\label{fig1}
\end{figure}

   We further take the simplest Gaussian profile $\gamma(r)=p_ie^{-r^2/d^2}$ (other bell-shaped gain profiles yield qualitatively similar results). 
Typical shapes of the simplest radially symmetric dissipative solitons  
obtained with the relaxation method are shown in Fig.~\ref{fig1}. For a given 
$\alpha$, the  soliton parameters (including $b$) are dictated by the gain coefficient. When $p_i$  is sufficiently small the solitons are broad [Fig.~\ref{fig1}(a)]. The field can expand far beyond the spatial region where the gain is realized due to the energy flow outward the amplifying domain.   Growth of $p_i$  is accompanied by progressively increasing localization [c.f.  panels (a) and (b) in Fig.~\ref{fig1}].  For   $\alpha\gtrsim 2$   with increase of $p_i$    the soliton shape takes on the form of a narrow peak superimposed on broader beam [Fig.~\ref{fig1} (b)].
 
The energy flow $U$  of a dissipative soliton is a nonmonotonic function of the gain coefficient $p_i$  [Fig.~\ref{fig2}(a)].   For small and moderate nonlinear losses the initial decrease of $U$  is followed by its growth, and then again replaced by decrease as $p_i$  increases [Fig.~\ref{fig2} (a)]. In the cutoff point $p_i^{co}$
the tangential line to $U(p)$   is vertical, but the value $U^{co}=U(p_i^{co})$ remains finite. We notice that it is possible to find the second branch of   solutions vanishing in the same point $(p_i^{co},U^{co})$   for which energy flow increases with  $p_i$ in the vicinity of cutoff,  {i.e. in $(p_i^{co},U^{co})$  two soliton families smoothly join, rather than one family terminates}. However, this second branch is unstable as the linear stability analysis predicts. The character of $U(p_i)$  dependence changes dramatically when $\alpha$ is sufficiently large [Fig.~\ref{fig2}(b)]. In this case $U$  grows with $p_i$   everywhere except for a region close to the cutoff. Interestingly, this change in the behavior of $U(p_i)$ accompanies the tendency for development of a narrow peak in the center of soliton. $b$ increases monotonically with $p_i$  for small and moderate   values of $\alpha$, but when $\alpha$  approaches  2 the dependence  $b(p_i)$  becomes nonmonotonic [Fig.~\ref{fig2}(c)].  {For large $\alpha$ one also observes   two soliton families merging at $p_i^{co}$.  
At a fixed width of the amplifying domain, $p_i^{co}$ grows almost linearly with increase of nonlinear losses [Fig.~\ref{fig2}(e)]. Meantime, $p_i^{co}$  diminishes with increase of $d$, so that at   $d\to\infty$ we naturally recover the case of medium with uniform gain where solitons can be found for any $p_i$ [Fig.~\ref{fig1}(f)]. Notice however, that for 2D solitons considered here $p_i^{cr}$ value also diminishes with $d$, so that while localized solitons at $d\to\infty$ exist for any $p_i$ they are all unstable.

\begin{figure}[h]
   \begin{tabular}{c}
\includegraphics[width=0.9\columnwidth]{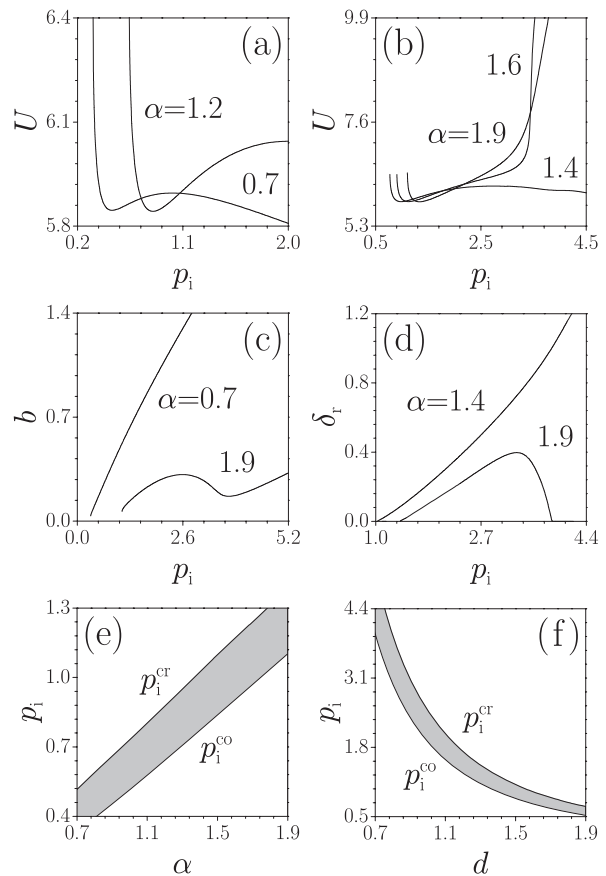}
   \end{tabular}
\caption{(Color online) Energy flow of radially symmetric dissipative soliton (a),(b), its propagation constant (c), and real part of perturbation growth rate   (d)   {\it vs} $p_i$   for different $\alpha$  values and $d=1.5$. Cutoff and critical values of gain coefficient (e) {\it vs} $\alpha$  at $d=1.5$  and  (f) {\it vs} $d$  at $\alpha=1.5$.}
\label{fig2}
\end{figure}

The stabilization of 2D solitons 
   means that for the respective initial conditions the localized gain does not induce collapse, which is  arrested by the nonlinear dissipation. On the other hand the nonlinear dissipation suppresses instability of the trivial zero solution preventing growth of the background noise. This is in   contrast to the case of Kerr medium with uniform gain and two-photon absorption where solitons are unstable even in 1D geometries \cite{Malomed2}. These facts have been numerically verified by the linear stability analysis with the ansatz 
$q=[w+v_1 e^{\delta\xi+in\varphi}+v_2^* e^{\delta^*\xi-in\varphi}]$, where $n$ is an integer azimuthal index. Solution of the respective linearized equations  reveals a surprising result:   most destructive perturbations correspond to $n=1$, rather than to $n=0$.
%, as it occurs for solitons in the conservative cubic medium. 
The typical dependencies of real part of perturbation growth rate $\delta_r$  on $p_i$   for $n=1$   are shown in Fig.~\ref{fig2}(d). For small and moderate nonlinear losses $\delta_r$  vanishes for any $n$  when $p_i^{co}<p_i<p_i^{cr}$, where $p_i^{cr}$ is some critical value. Thus the dissipative solitons are linearly stable in the region adjacent to the cutoff $p_i^{co}$.   The stability domain slowly expands with increase of $\alpha$ [Fig.~\ref{fig2}(e), shaded] and broadens considerably with decrease of the width $d$  of amplifying domain [Fig.~\ref{fig2}(f), shaded]. 
%Notice that stability domain extends a bit farther than the point where energy flow   acquires local minimum [see Fig.~\ref{fig2}(a)]. 
At small and moderate $\alpha$  inside the instability domain  $\delta_r$ increases monotonically with $p_i$, but for sufficiently high nonlinear losses the character of this dependence changes [Fig.~\ref{fig2}(d)] and another stability domain appears also at high $p_i$  values. This stability domain is limited since with further increase of  $p_i$ perturbations with $n=2$  become destructive. Results of linear stability analysis are supported by direct propagation of perturbed radially-symmetric dissipative solitons. Stable propagation of  a relatively broad soliton taken from stability domain adjacent to $p_i^{co}$   is shown in Fig.~\ref{fig3}(a), while stable propagation of a high-amplitude soliton with the shape resembling a narrow peak on broader pedestal from stability domain at high $p_i$  and $\alpha$  is shown in Fig.~\ref{fig3}(b).
\begin{figure}[h]
\begin{tabular}{c}
\includegraphics[width=0.9\columnwidth]{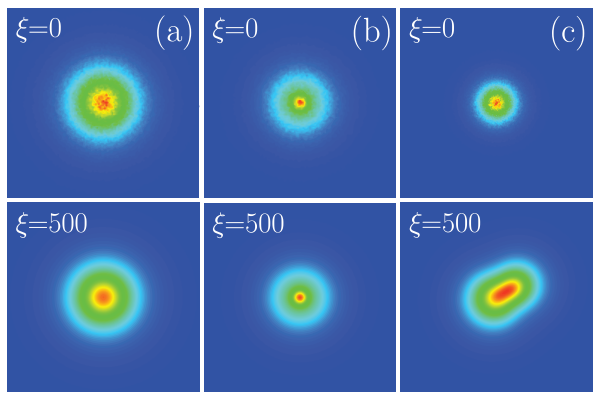}
\end{tabular}
\caption{(Color online) Stable propagation of radially-symmetric solitons at (a) $p_i=0.8$, $\alpha=1.2$, and (b) $p_i=4.2$,  $\alpha=1.9$, and (c) transformation of unstable radially-symmetric soliton into asymmetric stable state at $p_i=2.0$, $\alpha=1.2$. Field modulus distributions are shown at different distances. In all cases $d=1.5$.}
\label{fig3}
\end{figure}

The development of instability for perturbed unstable radially symmetric solitons results in their transformation into strongly asymmetric non-rotating dissipative soliton whose center is slightly displaced from the origin towards the periphery of amplifying domain [Fig.~\ref{fig3}(c)]. 
%Remarkably, 
Such states can be excited with Gaussian beams too, i.e. they are attractors with large basin.
%Such states can be excited not only upon decay of radially symmetric solitons, but also with beams %having arbitrary (e.g., Gaussian) shapes, i.e. they are attractors with a relatively large basin. 
The energy flow and propagation constant of the asymmetric soliton remain  the same as long as $p_i$, $d$, and $\alpha$   are fixed and only the beam orientation changes for different inputs and noise realizations. Thus the medium with localized gain and nonlinear losses can support strongly asymmetric dissipative solitons despite the fact that all parameters in the system are either uniform (diffraction, nonlinearity, and losses) or radially symmetric (gain). 
%This symmetry breaking is another central result of this Letter and it is in sharp contrast to conservative uniform  medium with Kerr-type nonlinearity that can not support stable stationary asymmetric states. 
The asymmetric solitons 
are characterized by a specific phase distribution [Figs.~\ref{fig4}(a) and (b)] that is asymmetric along longer axis of the beam and symmetric along its shorter axis.  
\begin{figure}[h]
\includegraphics[width=0.9\columnwidth]{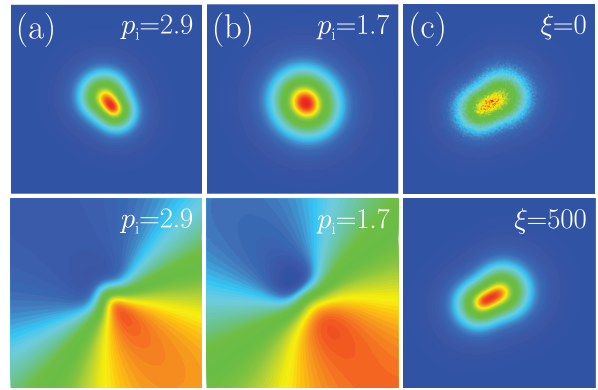}
\caption{(Color online) Field modulus (top) and phase (bottom) distributions in asymmetric dissipative solitons for $p_i=2.9$  (a) and  $p_i=1.7$ (b) at $\alpha=1.9$, $d=1.5$. (c) Stable propagation of a perturbed asymmetric soliton at $p_i=2.0$,  $\alpha=1.2$, $d=1.5$. Initial and final field modulus distributions are shown.}
\label{fig4}
\end{figure}
  
Using a method of direct propagation
we obtained the entire family of asymmetric solitons. The dependencies of energy flow and integral widths $d_{\eta}=2\left( \int\eta^2|q|^2d\br\right/U)^{1/2}$   ($d_\zeta$ is defined similarly)  on $p_i$ are shown in Fig.~\ref{fig5}. The family of asymmetric solitons bifurcates from the family of radially symmetric solitons with increase of $p_i$.  
The bifurcation occurs exactly in the point $p_i^{cr}$  where the radially symmetric solitons become unstable. With increase of  $p_i$ the asymmetry of soliton shape increases [c.f. Figs.~\ref{fig4}(a) and (b)]. The obtained asymmetric solitons propagate stably in the presence of small-scale input noise that also do not change the orientation of soliton in the transverse plane [Fig.~\ref{fig4}(c)].

\begin{figure}[h]
\includegraphics[width=0.9\columnwidth]{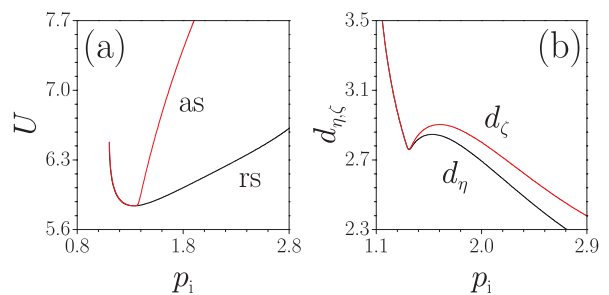}
\caption{(Color online) (a) Energy flows of asymmetric (red curve) and for radially symmetric (black curve) solitons  {\it vs} $p_i$  at $\alpha=1.9$, $d=1.5$.   (b)  $d_{\eta\zeta}$ {\it vs} $p_i$   for the family of asymmetric solitons whose representative shapes are shown in Figs.~\ref{fig4}(a) and (b).}
\label{fig5}
\end{figure}

Summarizing, we predicted that localized gain can support stable 2D 
solitons in a medium with Kerr-type nonlinearity and  strong two-photon absorption. This is in drastic contrast to behavior of solitons 
 {either in 2D conservative systems described by the nonlinear Schr\"odinger equation,}
 or in dissipative systems with uniform gain.  
     Moreover we obtained a stable family of asymmetric nonrotating solitons which have no counterparts in conservative or dissipative uniform systems with cubic nonlinearity.

\end{document}